\documentclass[aps,prd,superscriptaddress,twoside,twocolumn,nofootinbib,%
10pt,showpacs,floatfix]{revtex4-1}

\usepackage{amsmath,amssymb}
\usepackage{graphicx,bm}
\usepackage{ulem}
\usepackage{slashed}
\usepackage{dcolumn}
\usepackage{epsfig}
\usepackage{multirow}

\allowdisplaybreaks

\begin{document}


\title{Testing the tetraquark mixing framework from QCD sum rules for $a_0(980)$}


\author{Hee-Jung Lee}%
\email{hjl@chungbuk.ac.kr}
\affiliation{Department of Physics Education, Chungbuk National University, Cheongju, Chungbuk 28644, Korea}

\author{K. S. Kim}%
\affiliation{School of Liberal Arts and Science, Korea Aerospace University, Goyang, 412-791, Korea}

\author{Hungchong Kim}%
\email{hungchong@kau.ac.kr}
\affiliation{Research Institute of Basic Science, Korea Aerospace University, Goyang, 412-791, Korea}
\affiliation{Center for Extreme Nuclear Matters, Korea University, Seoul 02841, Korea}

\date{\today}


\begin{abstract}

According to a recent proposal of the tetraquark mixing framework,
the two light-meson nonets in the $J^{P}=0^{+}$ channel, namely
the light nonet composed of $a_0 (980)$, $K_0^* (800)$, $f_0 (500)$, $f_0(980)$,
and the heavy nonet of $a_0 (1450)$, $K_0^* (1430)$, $f_0 (1370)$, $f_0 (1500)$,
can be expressed by linear combinations of the two tetraquark types, one type containing the spin-0 diquark
and the other with the spin-1 diquark.
Among various consequences of this mixing model, one surprising result is that the second tetraquark with the spin-1
diquark configuration is more important for the light nonet.
In this work, we report that this result can be supported by the QCD sum rule calculation.
In particular, we construct a QCD sum rule for the isovector resonance $a_0(980)$ using an interpolating field
composed of both tetraquark types and then perform the operator product expansion up to dimension 10 operators.
Our sum rule analysis shows that the spin-1 diquark configuration
is crucial in generating the $a_0(980)$ mass.
Also, the mixed correlation function constructed from the two tetraquark types
is found to have large strength which seems consistent with
what the tetraquark mixing framework is advocating.
On the other hand, the correlation function from the interpolating field with
the spin-0 diquark configuration alone fails to predict the $a_0(980)$ mass
mostly by the huge negative contribution from dimension 8 operators.

\end{abstract}

\maketitle

\section{Introduction}
\label{sec:intro}

Tetraquarks have been anticipated for long time in hadron community.
Recently, with the development of high-energy facilities,
tetraquark candidates are accumulating from worldwide experiments in the heavy quark sector.
There are some candidates with charm quarks which include
the pioneering state $X(3872)$~\cite{Belle03,Aubert:2004zr, Choi:2011fc, Aaij:2013zoa}
measured in the $B$-meson decays as well as other similar states
$X(3823)$, $X(3900)$, $X(3940)$, $X(4140)$, $X(4274)$, $X(4500)$,
$X(4700)$~\cite{Bhardwaj:2013rmw,Xiao:2013iha,Abe:2007jna, Aaij:2016iza,Aaij:2016nsc}.
These states have been investigated theoretically with their possibility of being
hidden-charm tetraquarks~\cite{Maiani:2004vq,Kim:2016tys,Anwar:2018sol,Zhao:2014qva}.
Possibility for open-bottom and open-charm tetraquarks has been
investigated in Ref.~\cite{Kim:2014ywa} for the resonances which are normally
treated as the $B$-, $D$-meson excited states.

Also tetraquarks are expected to exist in the light quark sector composed of $u,d,s$ quarks.
In fact, as is well known, the tetraquark study in the light quark system
can be traced back to 1970s when
Jaffe proposed a fascinating model of diquark-antidiquark~\cite{Jaffe77a,Jaffe77b,Jaffe04}.
In this model, tetraquarks are constructed by combining the spin-0 diquark,
in the color and flavor structures of ($\bar{\bm{3}}_c, \bar{\bm{3}}_f$),
with its antidiquark in ($\bm{3}_c, \bm{3}_f$) so that the resulting tetraquarks form a flavor nonet
($\bar{\bm{3}}_f\otimes \bm{3}_f = \bm{1}_f\oplus \bm{8}_f$).
The spin-0 diquark is adopted because it is most attractive among all the possible diquarks if the binding is calculated from
the color-spin interaction~\cite{Jaffe:1999ze}.
Thus, it is commonly expected that the resulting tetraquarks are stable.

But one may ask whether the spin-0 diquark (and the corresponding antidiquark)
is the only building block to construct stable tetraquarks under the diquark-antidiquark approach.
Since the total binding energy is calculated by summing over pairwise interactions among all the constituting quarks,
the diquark binding may not be the sole criterion in judging stable tetraquarks.
To be specific, the diquark (and antidiquark) binding constitutes only the part of the total binding energy.
There are additional contributions from other pairs like quark-antiquark.
In this sense, it is necessary to examine other diquarks in addition to the spin-0 diquark
as possible constituents in making stable tetraquarks.
Indeed, as recently advocated by Refs.~\cite{Kim:2016dfq,Kim:2017yur,Kim:2017yvd,Kim:2018zob}, one can construct
the second tetraquark by using the spin-1 diquark with the color and flavor structure ($\bm{6}_c, \bar{\bm{3}}_f$).
Even though the spin-1 diquark is less compact than the spin-0 diquark, the total binding energy
of the second tetraquark,
if calculated from the color-spin interaction for all the pairs, is found to
be more attractive than the binding from the first tetraquark type involving the spin-0 diquark.
Therefore, two types are possible for the stable tetraquark.
Having the same flavor structure as the spin-0 diquark, the tetraquarks with the spin-1 diquark configuration also form a flavor nonet.
Both tetraquarks have the same quantum numbers $J^{P}=0^{+}$ by their construction.

What is interesting is that the two tetraquark types mix strongly each other through the color-spin interaction.
Two eigenstates that diagonalize the color-spin interaction
can be identified by physical resonances because
they also diagonalize the other terms in the Hamiltonian, the color-electric potential as well as the quark mass terms.
In other words, the physical states are linear combinations of the two tetraquark types.
This tetraquark mixing framework seems to explain very much the two nonets
that can be found in Particle Data Group(PDG)~\cite{PDG18} in the $J^{P}=0^{+}$ channel, namely
the light nonet composed of the lowest-lying resonances, $a_0 (980)$, $K_0^* (800)$, $f_0 (500)$, $f_0(980)$, and
the heavy nonet whose members are the next higher resonances, $a_0 (1450)$, $K_0^* (1430)$, $f_0 (1370)$, $f_0 (1500)$.
These two nonets are well separated in mass from the rest resonances in PDG.
In fact, Ref.~\cite{Kim:2018zob} presented various phenomenological
signatures to support this mixing framework including not only the famous inverted mass spectrum among the nonet members
but also the others related to the hyperfine mass splittings,
the mixing parameters, the Gell-Mann--Okubo mass relation, the enhancement or suppression of
the fall-apart decay modes and so on.
Therefore it is quite promising that this mixing scheme is indeed realized by the two nonets in PDG.

In order to solidify this picture further, it may be desirable to test various consequences
from dynamical calculations based on the fundamental theory like quantum chromodynamics (QCD).
In practice, the QCD sum rule~\cite{Shifman:1978bx,Shifman:1978by,Reinders:1984sr}
or lattice QCD calculation~\cite{Dudek:2016cru} can be used for this purpose.
One surprising result to test is the statement that the light nonet
has more probability to stay in the second type tetraquark containing the spin-1 diquark
rather than in the first type involving the spin-0 diquark~\cite{Kim:2016dfq,Kim:2017yur,Kim:2017yvd,Kim:2018zob}.
This picture is very different from the common expectation that the light nonet
has the structure of the first tetraquark type only~\cite{Jaffe77a,Jaffe77b,Jaffe04,Jaffe:1999ze,EFG09,Santopinto:2006my}.

In this regard, it will be particularly interesting to revisit
the QCD sum rule calculation performed by one of the present authors (HJL) in Ref.~\cite{Lee:2005hs}.
There, the QCD sum rule is constructed by using an interpolating field based on the first tetraquark type
only but the result is not conclusive in extracting the light nonet mass mainly because
of the huge negative contribution from dimension 8 operators. This may indicate that
the first tetraquark type does not represent the light nonet properly.
The similar result is reported also by the later calculations~\cite{Wang:2015uha}.
In our point of view, the failure of this QCD sum rule may be closely related to
the statement above that the spin-0 diquark configuration is less probable for the light nonet.
Instead, the spin-1 diquark configuration may be more important for the light nonet.
Therefore it may be worth performing the QCD sum rule calculation again but with an interpolating field
incorporating both spin-0 and spin-1 diquark configurations.

In this work, we present a QCD sum rule study for the light nonet in order to test the tetraquark mixing framework.
In this study, we take the isovector resonance $a_0(980)$ among the light nonet members and this choice will
be justified in Sec.~\ref{sec:review}.
An interpolating field for $a_0(980)$ will be constructed in Sec.~\ref{sec:interpolating field} based on the tetraquark mixing framework.
Then using this interpolating field, we construct the corresponding QCD sum rule in Sec.~\ref{sec:QCDSR}
by performing the operator product expansion (OPE) up to dimension 10. The results will be discussed in Sec.~\ref{sec:result}.
We summarize in Sec.~\ref{sec:summary}.

\section{Tetraquark mixing framework}
\label{sec:review}

To motivate the construction of an interpolating field in our QCD sum rule study,
we briefly look at the mathematical structure of the tetraquark mixing framework
advocated by Refs.~\cite{Kim:2016dfq,Kim:2017yur,Kim:2017yvd,Kim:2018zob}.
The mixing framework has been developed as a possible structure for the
two nonets in PDG, the light nonet composed of $a_0 (980)$, $K_0^* (800)$, $f_0 (500)$, $f_0(980)$,
and the heavy nonet of $a_0 (1450)$, $K_0^* (1430)$, $f_0 (1370)$, $f_0 (1500)$.
According to this mixing framework, one can introduce two types of tetraquark in the diquark-antidiquark model.
The first tetraquark type, which is commonly adopted in the tetraquark studies,
is constructed by combining the spin-0 diquark, whose color and flavor structures are
in ($\bar{\bf 3}_c, \bar{\bf 3}_f$),
and the corresponding spin-0 antidiquark~\cite{Jaffe77a,Jaffe77b,Jaffe04}. This first tetraquark type is denoted by
$| 000 \rangle$ where the first number represents the tetraquark spin,
the second the diquark spin, and the third the antidiquark spin.
The second tetraquark type, $| 011\rangle$, which was suggested as another possibility in
Refs.~\cite{Kim:2016dfq,Kim:2017yur,Kim:2017yvd,Kim:2018zob,Black:1998wt},
is constructed by combining the spin-1 diquark in the structure of (${\bf 6}_c, \bar{\bf 3}_f$) and its antidiquark.
By construction, the two tetraquark types differ by color and spin configurations
but they have the same flavor structure, namely the nonet.

The color structure of the two tetraquarks can be explicitly written as~\footnote{See Ref.~\cite{Oh:2004gz}
for technical details in using a tensor notation for SU(3).}
\begin{eqnarray}
&&|000\rangle:\ \frac{1}{\sqrt{12}}  \varepsilon_{abd}^{} \ \varepsilon_{aef}
\left ( q^b q^d \right )
\left ( \bar{q}^e \bar{q}^f \right )\label{color1}\ ,\\
&&|011\rangle:\ \frac{1}{\sqrt{96}} \left ( q^a q^b+q^b q^a \right )
\left (\bar{q}^a \bar{q}^b+\bar{q}^b \bar{q}^a\right )\label{color2}\ ,
\end{eqnarray}
where the Roman indices, $a,b,d,e,f$, denote the colors.
Both tetraquarks form the same flavor nonet which can be broken down
to an octet and a singlet. Their members in tensor notation can be expressed by
\begin{eqnarray}
[{\bf 8}_f]^i_{j} &=& T_{j}\bar{T}^{i}-\frac{1}{3} \delta^{i}_{j}~T_{m}\bar{T}^{m}\label{octet}\ ,\\
{\bf 1}_f &=& \frac{1}{\sqrt{3}}T_{m}\bar{T}^{m}\label{singlet}\ .
\end{eqnarray}
Here $T_i$ ($\bar{T}^i$) denotes the diquark (antidiquark) defined by
\begin{eqnarray}
T^i &=&\frac{1}{\sqrt{2}}\epsilon^{ijk}q_j q_k\equiv [q_j q_k]\ ,\nonumber\\
\bar{T}_i &=& \frac{1}{\sqrt{2}}\epsilon_{ijk}\bar{q}^j \bar{q}^k \equiv [{\bar q}^j {\bar q}^k] \ ,
\end{eqnarray}
with respect to the quark flavors, $q_i=u,d,s$ ($\bar{q}^i=\bar{u},\bar{d},\bar{s}$).

The most striking feature is that the two tetraquark types mix each other strongly through the color-spin
interaction and the two nonets in PDG
can be collectively represented by linear combinations of the two tetraquark types as
\begin{eqnarray}
|\text{heavy~nonet} \rangle &=& -\alpha | 000 \rangle + \beta |011 \rangle\label{heavy}\ ,\\
|\text{light~nonet} \rangle~ &=&\beta | 000 \rangle + \alpha |011 \rangle\label{light}\ ,
\end{eqnarray}
which diagonalize the color-spin interaction.
Here $\alpha, \beta$ are the mixing parameters determined in each isospin channel by
the diagonalization process. But they are found to be almost independent of
isospin~\cite{Kim:2016dfq,Kim:2017yur,Kim:2017yvd,Kim:2018zob} and their values are approximately close to
$\alpha\approx\sqrt{2/3}, \beta\approx\sqrt{1/3}$. This basically implies that the wave functions in
Eqs.~(\ref{heavy}), (\ref{light}) separately form an approximate flavor nonet
consequently supporting the identification in terms of the two nonets in PDG.

One surprising result for the mixing parameters is the inequality, $\alpha > \beta$.
As one can see in Eq.~(\ref{light}), this inequality
means that the light nonet members,
$a_0 (980)$, $K_0^* (800)$, $f_0 (500)$, $f_0(980)$, are more dominated by the spin-1 diquark
configuration rather than the spin-0 diquark
configuration.  To be specific, the probability to stay in the first tetraquark
type is about 33\% and that in the second type is 67\%.
Therefore the spin-1 diquark configuration is more important in the light nonet members
when they are described by tetraquarks.
We stress again that this picture is very different from the common expectation that the light nonet members
are dominated by the spin-0 diquark configuration.

Our primary task is to test this surprising
result in the light nonet by QCD sum rules~\cite{Shifman:1978bx,Shifman:1978by,Reinders:1984sr}.
In principle, any member in the light nonet can be tested for our purpose but in practice
some care needs to be taken in choosing one specific resonance to work on.
In this work, we choose the isovector resonance, $a_0(980)$, because, first of all,
this is a relatively sharp resonance with small decay width.
So the pole and continuum ansatz, which is the inevitable prescription in QCD sum rules, may fit better
to $a_0(980)$ than the other broad resonances like $K^*_0(800), f_0 (500)$.  Moreover, the $a_0(980)$ resonance has another advantage
over the isoscalar members $f_0 (500)$, $f_0(980)$ because $a_0(980)$ does not
suffer from additional ambiguity coming from the flavor mixing~\cite{Kim:2017yvd} which can be referred as
the generalized Okubo-Zweig-Iizuka (OZI) rule.

\section{Interpolating field for $a_0(980)$}
\label{sec:interpolating field}

To investigate $a_0(980)$ through QCD sum rules, we need to construct an interpolating field for $a_0(980)$
that properly incorporates the tetraquark mixing framework developed in the constituent quark picture.
The mixing framework suggests that the isovector resonance
$a_0(980)$ is represented by the mixture of the two tetraquark types in the $I=1$ channel,
\begin{equation}
|a_0(980)\rangle=\beta|000\rangle_{I=1} + \alpha|011\rangle_{I=1}\ .
\label{a0wave}
\end{equation}
The mixing parameters in this $I=1$ channel are determined to be~\cite{Kim:2016dfq,Kim:2017yur,Kim:2017yvd,Kim:2018zob}
\begin{equation}
\alpha=0.8167\ ,\quad \beta=0.5770\label{mixing parameters}\ .
\end{equation}
The flavor structure of $a_0(980)$, which is the $[{\bf 8}_f]^1_{2}$ member in Eq.~(\ref{octet})~\footnote{Among isovector members,
we choose the charged member $a^+_0(980)$ in this study.}, takes the form
\begin{equation}
[su][\bar{d}\bar{s}]=\frac{1}{\sqrt{2}}(su-us)\frac{1}{\sqrt{2}}(\bar{d}\bar{s}-\bar{s}\bar{d})\ .
\end{equation}
The color structure of $[su][\bar{d}\bar{s}]$ is given by Eq.~(\ref{color1}) for the $|000\rangle$
case and by Eq.~(\ref{color2}) for the $|011\rangle$
case.  It should be remembered that this structure for $a_0(980)$ is based on the constituent quark picture having
all the quarks in an $S$-wave.

To construct an interpolating field with {\it current quarks}, we need to replace the constituting
quarks by the Dirac spinors while keeping
the color and flavor structures as above.
Then the remaining task is to determine appropriate Dirac
structures to be inserted between the two quarks for the spin-0 diquark as well as for spin-1 diquark.
One more thing to be kept in mind is that the interpolating field must be nonzero
and should be normalized as $|000\rangle_{I=1}$, $|011\rangle_{I=1}$ in the static limit so that
one can facilitate the same mixing parameters as given in Eq.~(\ref{mixing parameters}).

To start, we write down general forms of the interpolating fields for the two diquark types,
one for the spin-0 diquark with the color and flavor structures ($\bar{\bf 3}_c, \bar{\bf 3}_f$),
and the other for spin-1 diquark with (${\bf 6}_c, \bar{\bf 3}_f$). That is, for the member containing $u,s$ quarks,
\begin{eqnarray}
\text{spin-0:}~&&\epsilon_{abc}(s_b^T\Gamma_0 u_c-u_b^T\Gamma_0 s_c)\label{spin0}\ ,
\\
\text{spin-1:}~&& s^T_a\Gamma_1u_b+s^T_b\Gamma_1u_a-u^T_a\Gamma_1s_b-u^T_b\Gamma_1s_a\label{spin1}\ .
\end{eqnarray}
The $4\times 4$ matrices, $\Gamma_{0,1}$, whose subscript denotes the associated diquark,
can be fixed as follows.

A standard way to construct a diquark interpolating field
is to replace the antiquark $\bar{q}$ in the mesonic field of the form $\bar{q}\Gamma q$
by its charge conjugation analog, $\bar{q}\rightarrow q^TC$~\cite{griegel}. Then $\Gamma_{0}$ in Eq.~(\ref{spin0})
takes the form $C\Gamma$ where $\Gamma$ is a Dirac matrix to be chosen from
the 16 independent matrices, $1,\gamma_5, \gamma_\mu, \gamma_5 \gamma_\mu, \sigma_{\mu\nu}$.
Using the basic properties among Dirac matrices, one can easily prove that
\begin{equation}
\Gamma_{0}^T=\pm \Gamma_{0}\label{constraint}
\end{equation}
for all the possible $\Gamma$.
To take advantage of this identity, we rewrite the second term in Eq.~(\ref{spin0}) as
\begin{eqnarray}
\epsilon_{abc}u_b^T\Gamma_0 s_c
=-\epsilon_{abc}s_c^T\Gamma_0^T u_b
=\epsilon_{abc}s_b^T\Gamma_0^T u_c\ ,
\end{eqnarray}
where the anticommutation relation $\{u,s\}=0$ has been used in the first step.
Then the two terms in Eq.~(\ref{spin0}) can be combined into
\begin{equation}
\epsilon_{abc}s_b^T(\Gamma_0 - \Gamma_0^T) u_c\label{combine}\ .
\end{equation}
In order for this diquark to be nonzero, we should have
$\Gamma_{0}^T=- \Gamma_{0}$ among the two possibilities in Eq.~(\ref{constraint}).
This condition, if imposed on the spin-0 diquark, leads to $\Gamma_0=C,~C\gamma_5$.~\footnote{These structures turn out
to be the same as determined simply from the diquark spin being zero. But our prescription becomes more restrictive
when it determines the structure of the spin-1 diquark.}
Since the charge conjugation $C$
is off-diagonal in Dirac space, it connects the upper and lower components of the Dirac spinors
when it is plugged into Eq.~(\ref{combine}). Thus, this diquark with $\Gamma_0=C$ vanishes in the static limit indicating
that this diquark is not relevant.
Instead, the diquark with $\Gamma_0=C\gamma_5$ does not vanish in the static limit and we can take this
as the appropriate diquark in this spin-0 case.

The similar steps can be taken for the spin-1 diquark, Eq.~(\ref{spin1}), and in this case, we find the different
constraint, $\Gamma_1^T=\Gamma_1$, essentially due to that the diquark in Eq.~(\ref{spin1})
does not entail $\epsilon_{abc}$.
This constraint, if imposed on the spin-1 diquark, leads to $\Gamma_1=C\gamma_\mu$.
In summary, we come up with the following Dirac structures
\begin{eqnarray}
\Gamma_0=C\gamma_5\ ,\quad \Gamma_1=C\gamma_\mu\ ,
\end{eqnarray}
as the relevant ones for the spin-0 and spin-1 diquark, respectively.

Now, combining with the corresponding antidiquarks, and after some minor manipulations, we obtain
the interpolating fields for the two tetraquark types as
\begin{eqnarray}
J_{0}&=&\frac{1}{\sqrt{12}}\epsilon_{abc}\epsilon_{ade}(s_b^T\Gamma_0 u_c)(\bar{d}_d\tilde{\Gamma}_0\bar{s}_e^T)\label{type0}\ ,\\
J_{1}&=&\frac{1}{\sqrt{72}}(s^T_a\Gamma_1u_b)
(\bar{d}_a\tilde{\Gamma}_1\bar{s}^T_b+\bar{d}_b\tilde{\Gamma}_1\bar{s}^T_a)\label{type1}\ ,
\end{eqnarray}
where $\tilde{\Gamma}_{0,1}=\gamma^0\Gamma_{0,1}^\dagger\gamma^0$.
Again, the subscript in $J_0,J_1$ denotes the diquark type involved.
The numerical factors in front of these equations
are chosen to make the interpolating fields reproduce the same normalized states in the static limit.
Note also that the Lorentz indices in $J_{1}$ should be contracted in order to make the spin-0 tetraquark state.

Finally, the interpolating field for the light-nonet isovector member, $a_0(980)$, can be
expressed by a linear combination of $J_0, J_1$ similarly to its static correspondence of
Eq.~(\ref{a0wave}). That is, the interpolating field for $a_0(980)$ can be written as
\begin{equation}
J^L_{a_0}=\beta J_{0}+\alpha J_{1}\label{if}\ ,
\end{equation}
where the superscript ``$L$'' has been introduced to denote the light nonet member.
Of course, the interpolating field containing the spin-0 diquark, $J_0$, is not new and this type has been often used elsewhere
for investigating tetraquark possibility in QCD sum rules~\cite{Lee:2005hs,Agaev:2018fvz,Kim:2005gt}.
But the second type $J_1$, which involves the spin-1 diquark, is not conventional in the study of the light nonet
in terms of tetraquarks.
Our main goal in this work is to
investigate the role of this additional component as well as its mixing with $J_0$ from QCD sum rules.
There are of course different types of tetraquark interpolating fields like the ones
introduced in Refs.~\cite{Kojo:2008hk,Chen:2007xr,Chen:2006zh}
whose connection to the teraquark mixing framework is unclear at the moment.

\section{QCD sum rule for $a_0(980)$}
\label{sec:QCDSR}

In this section, we illustrate how we construct a QCD sum rule for $a_0(980)$
using the interpolating field developed in Sec.~\ref{sec:interpolating field}.
In this sum rule, we consider the following correlation function
\begin{equation}
\Pi(q^2)=i\int d^4x e^{iq\cdot x}
\langle0|TJ^L_{a_0}(x)J^{L\dagger}_{a_0}(0)|0\rangle\label{correlator}\ ,
\end{equation}
with the interpolating field $J^L_{a_0}$ given by Eq.~(\ref{if}).
As usually done in QCD sum rules,
this correlation function is evaluated in two ways. In the one hand, the operator product
expansion (OPE) is performed to express the correlator in terms of QCD degrees of freedom.
In practice, the OPE calculation is truncated up to certain dimension because the OPE is
expected to converge as the operator dimension grows.
Its validity therefore relies on whether high dimensional operators contribute less to the truncated OPE.
In the other hand, the phenomenological ansatz for the correlator is constructed using
hadronic degrees of freedom. This ansatz involves the lowest-lying state of concern and higher resonances
which are normally treated as the continuum modeled by the QCD duality assumption.

Through a dispersion relation, the correlation function $\Pi(q^2)$ can be expressed by its spectral density via
\begin{equation}
\Pi^{\rm OPE, phen}(q^2)=\frac{1}{\pi}\int_0^\infty
ds~\frac{{\rm Im}\Pi^{\rm OPE, phen}(s)}{s-q^2}\label{sr}\ ,
\end{equation}
for the OPE and phenomenological side respectively.
The two sides are then matched after the Borel transformation, which eventually leads to the familiar sum rule
equation relating the two spectral densities,
\begin{eqnarray}
\int_0^{s_0} ds \frac{1}{\pi}{\rm Im}\!\left
[\Pi^{\rm OPE}(s)-\Pi^{\rm phen}(s)\right ]e^{-s/M^2}=0\label{sumrule1}\ ,
\end{eqnarray}
where $s_0, M$ denote the continuum threshold and the Borel mass respectively.
Through the QCD duality assumption,
higher resonance contributions in the phenomenological side are equated to the logarithmic cut in
the OPE above the continuum threshold $s_0$ so that the integral is
restricted to the interval $0\sim s_0$.
In addition, the Borel weight, $e^{-s/M^2}$, amplifies the lowest-lying pole contribution while suppressing higher
resonance contributions in the phenomenological side. Also this weight suppresses
high dimensional operators in the OPE side as it transforms the nonperturbative terms of the form $1/s^n, (n\ge 2)$
into $1/[(n-1)!(M^2)^{n-1}]$. Therefore, it is expected that Eq.~(\ref{sumrule1}) can be used to predict
some properties of the lowest-lying pole from the truncated OPE.
All these prescriptions, however, as they are being rough,
suggest that the results from QCD sum rules may not coincide precisely with
the hadronic parameters to be extracted. Instead, the results can
be regarded as qualitative guides.

Nevertheless, by following the prescriptions above, only the lowest-lying resonance
contributes to the phenomenological side.
Using the convention
\begin{equation}
\langle 0|J_{a_0}^L(0)|a_0^+\rangle=\sqrt{2}f_{a_0}m_{a_0}^4\label{coup}\ ,
\end{equation}
we get the phenomenological side of Eq.~(\ref{sumrule1}),
\begin{equation}
\int_0^{s_0} ds\frac{1}{\pi}{\rm Im}\Pi^{\rm Phen}(s)e^{-s/M^2}=2f_{a_0}^2m_{a_0}^8e^{-m_{a_0}^2/M^2}\label{phen}\ .
\end{equation}
The Borel-weighted integral of the OPE spectral density, which we denote by $\hat{\Pi}^{\rm OPE}(M^2)$,
can be divided into three parts
depending on the interpolating fields in Eqs.~(\ref{type0}),~(\ref{type1}).
Specifically, we have, for the OPE side of Eq.~(\ref{sumrule1}),
\begin{eqnarray}
\hat{\Pi}^{\rm OPE}(M^2)&\equiv&\int_0^{s_0} ds~ \frac{1}{\pi}{\rm Im}\Pi^{\rm OPE}(s)e^{-s/M^2}\nonumber\\
&=&\beta^2\hat{\Pi}^{\rm OPE}_{0,0}+2\beta\alpha\hat{\Pi}^{\rm OPE}_{0,1}
+\alpha^2\hat{\Pi}^{\rm OPE}_{1,1}\label{ope}\ ,
\end{eqnarray}
where the subscripts in the 2nd equation specify the interpolating fields, $J_0,J_1$, that participate
in this equation through Eq.~(\ref{if}).
Each correlator can be calculated straightforwardly~\footnote{We take the same technical steps as
in Refs.~\cite{Lee:2005hs,Lee:2005ny} in calculating the OPE expressions. So one may take
a look at these references for additional details.
One slight difference is the notation for the Dirac matrix $\sigma^{\mu\nu}$.
There, it was defined as $\sigma^{\mu\nu}\equiv \frac{1}{2}[\gamma^\mu,\gamma^\nu]$ while here we define it with
imaginary ``$i$'' so that $\sigma^{\mu\nu}\equiv \frac{i}{2}[\gamma^\mu,\gamma^\nu]$.}.
We obtain the OPE expressions for the three correlators up to dimension 10, after the Borel transformation, as
\begin{widetext}
\begin{eqnarray}
\hat{\Pi}^{\rm OPE}_{0,0}&=&\frac{1}{12}\Bigg\{\frac{M^{10}E_4(M^2)}{2^9\cdot5\pi^6}
+\frac{\langle g_c^2G^2\rangle}{2^{10}\cdot3\pi^6}M^6E_2(M^2) +\frac{m_s\left[\langle\bar{s}s\rangle
-2\langle\bar{q}q\rangle\right]}{2^5\cdot3\pi^4}M^6E_2(M^2)
\nonumber\\
&&
+\frac{\langle\bar{q}q\rangle\langle\bar{s}s\rangle}{2^2\cdot3\pi^2}M^4E_1(M^2)
+\frac{m_s [\langle\bar{s}g_c\sigma\cdot Gs\rangle+6\langle\bar{q}g_c\sigma\cdot Gq\rangle]}
{2^7\cdot3\pi^4}M^4E_1(M^2)
\nonumber\\
&&+\frac{m_s\langle\bar{q}g_c\sigma\cdot Gq\rangle}{2^6\pi^4}M^4\tilde{W}_1(M^2)
-\frac{1}{2^3\cdot3\pi^2}\left[\langle\bar{q}q\rangle \langle\bar{s}g_c\sigma\cdot Gs\rangle
+\langle\bar{s}s\rangle \langle\bar{q}g_c\sigma\cdot Gq\rangle\right]M^2E_0(M^2)
\nonumber\\
&&-\frac{m_s\langle g_c^2 G^2\rangle}{2^7\cdot3^2\pi^4}
\left[5\langle\bar{q}q\rangle-\frac{3}{2}\langle\bar{s}s\rangle\right]M^2E_0(M^2)
-\frac{m_s\langle g_c^2 G^2\rangle\langle\bar{q}q\rangle}{2^6\cdot3\pi^4}
M^2W_0(M^2)
\nonumber\\
&&
+\frac{59}{2^{9}\cdot3^2\pi^2 }\langle\bar{q}g_c\sigma\cdot Gq\rangle
\langle\bar{s}g_c\sigma\cdot Gs\rangle
+\frac{7\langle g_c^2 G^2\rangle\langle\bar{q}q\rangle\langle\bar{s}s\rangle}{2^5\cdot3^3\pi^2}
-\frac{m_s\langle\bar{q}q\rangle\langle\bar{s}s\rangle}{3^2}
\left[2\langle\bar{q}q\rangle-\langle\bar{s}s\rangle\right]\Bigg\}\label{ope00}\ ,
\\
\nonumber\\
\hat{\Pi}^{\rm OPE}_{0,1}
&=&\frac{1}{12\sqrt{6}}\Bigg\{-\frac{3m_s}{2^8\pi^4}\left[\langle\bar{q}g_c\sigma\cdot Gq\rangle
+\langle\bar{s}g_c\sigma\cdot Gs\rangle\right]M^4\left [E_1(M^2)-\tilde{W}_1(M^2)\right]
\nonumber\\
&&-\frac{m_s \langle g_c^2 G^2\rangle}{2^9\pi^4}
\left[\langle\bar{q}q\rangle-3\langle\bar{s}s\rangle\right]M^2E_0(M^2)
-\frac{m_s \langle g_c^2G^2\rangle}{2^9\pi^4}
\left[3\langle\bar{q}q\rangle-\langle\bar{s}s\rangle\right]M^2\tilde{W}_0(M^2)
\nonumber\\
&&+\frac{1}{2^5\pi^2}\left[\langle\bar{q}g_c\sigma\cdot Gq\rangle
+\langle\bar{s}g_c\sigma\cdot Gs\rangle\right]
\left[\langle\bar{q}q\rangle+\langle\bar{s}s\rangle\right]M^2E_0(M^2)
\nonumber\\
&&-\frac{17}{2^{10}\cdot3\pi^2}\left[\langle\bar{q}g_c\sigma\cdot Gq\rangle
+\langle\bar{s}g_c\sigma\cdot Gs\rangle\right]^2
-\frac{\langle g_c^2G^2\rangle}{2^6\cdot3\pi^2}
\left[\langle\bar{q}q\rangle^2+\langle\bar{s}s\rangle^2\right]\Bigg\}\label{ope01}\ ,
\\
\nonumber\\
\hat{\Pi}^{\rm OPE}_{1,1}&=&\frac{1}{72}\Bigg\{\frac{M^{10}E_4(M^2)}{2^6\cdot5\pi^6}
+\frac{5\langle g_c^2 G^2\rangle}{2^{9}\cdot3\pi^6}M^6E_2(M^2)+\frac{m_s\left[\langle\bar{s}s\rangle-
\langle\bar{q}q\rangle\right]}{2^2\cdot3\pi^4}M^6E_2(M^2)
\nonumber\\
&&
+\frac{\langle\bar{q}q\rangle\langle\bar{s}s\rangle}{3\pi^2}M^4E_1(M^2)
+\frac{m_s \left[9\langle\bar{q}g_c\sigma\cdot Gq\rangle+46\langle\bar{s}g_c\sigma\cdot Gs\rangle\right]}
{2^7\cdot3\pi^4}M^4E_1(M^2)
\nonumber\\
&&+\frac{1}{2^3\cdot3\pi^2}\left[\langle\bar{q}q\rangle \langle\bar{s}g_c\sigma\cdot Gs\rangle
-\frac{3}{2}\langle\bar{s}s\rangle \langle\bar{q} g_c\sigma\cdot Gq\rangle\right]M^2E_0(M^2)
\nonumber\\
&&-\frac{13m_s\langle g_c^2 G^2\rangle\langle\bar{q}q\rangle}{2^8\cdot3^2\pi^4}M^2E_0(M^2)
+\frac{m_s\langle g_c^2 G^2\rangle\langle\bar{s}s\rangle}{2^7\cdot3\pi^4}M^2E_0(M^2)
\nonumber\\
&&-\frac{m_s\langle g_c^2 G^2\rangle\langle\bar{q}q\rangle}{2^4\cdot3\pi^4}M^2W_0(M^2)
+\frac{5m_s\langle g_c^2 G^2\rangle\langle\bar{q}q\rangle}{2^6\cdot3\pi^4}M^2\tilde{W}_0(M^2)
\nonumber\\
&&+\frac{5}{2^{8}\pi^2 }\langle\bar{q}g_c\sigma\cdot Gq\rangle
\langle\bar{s}g_c\sigma\cdot Gs\rangle
-\frac{\langle g_c^2 G^2\rangle\langle\bar{q}q\rangle\langle\bar{s}s\rangle}{2^6\cdot3^3\pi^2}
-\frac{4m_s\langle\bar{q}q\rangle\langle\bar{s}s\rangle}{3^2}
\left[4\langle\bar{q}q\rangle-\langle\bar{s}s\rangle\right]\Bigg\}\label{ope11}\ .
\end{eqnarray}
\end{widetext}
Here $E_n(M^2), \tilde{W}_n(M^2), W_n(M^2)$ are the functions associated with the continuum threshold and their explicit
expression can be found in Refs~\cite{Lee:2005hs,Lee:2005ny}.  The overall factors, which are written outside of the brackets,
come from the normalizations in Eqs.~(\ref{type0}),(\ref{type1}).
Note also that the vacuum saturation hypothesis has been used in factorizing high dimensional operators
into lower dimensional ones.
In our numerical analysis, we use the conventional QCD parameters,
\begin{eqnarray}
&&\langle\bar{q}q\rangle=(-0.25)^3~{\rm GeV}^3~ \text{for}~ q=u,d~(\bar{q}=\bar{u},\bar{d}),
\nonumber\\
&&\langle\bar{q}g_c\sigma\cdot Gq\rangle=m_0^2\langle\bar{q}q\rangle
=0.8\langle\bar{q}q\rangle,
\nonumber\\
&&
\frac{\langle\bar{s}s\rangle}{\langle\bar{q}q\rangle}=
\frac{\langle\bar{s}g_c\sigma\cdot Gs\rangle}{\langle\bar{q}g_c\sigma\cdot Gq\rangle}=0.8, m_s=0.15~{\rm GeV},
\nonumber\\
&&\langle g_c^2 G^2\rangle=0.47~{\rm GeV}^4,
\Lambda=0.5~{\rm GeV}\label{QCDparameters} .
\end{eqnarray}

We rewrite the QCD sum rule for $a_0(980)$ succinctly as
\begin{equation}
2f_{a_0}^2m_{a_0}^8e^{-m_{a_0}^2/M^2}=\hat{\Pi}^{\rm OPE}(M^2)\label{final}\ ,
\end{equation}
where the phenomenological side has been taken from Eq.~(\ref{phen}) and the OPE side from Eq.~(\ref{ope}).
This final sum rule, Eq.~(\ref{final}), is a general formula
in a sense that it can be used also for the sum rule only with $J_0$ by setting the mixing parameters, $\alpha=0,\beta=1$,
and for the $J_1$ sum rule by $\alpha=1,\beta=0$. From these separate sum rules,
one can investigate the relative importance of the spin-0 and spin-1 diquark configurations in describing $a_0(980)$.
Another thing to mention is that the left hand side of Eq.~(\ref{final}) is positive definite, which can be
utilized as another
constraint~\cite{Lee:2005hs,Kim:2005gt} in testing the reliability of our QCD sum rules.

If the equation Eq.~(\ref{final}) is exact, then, as one can guess from the mathematical form of the left-hand side,
the $a_0(980)$ mass can be extracted from the corresponding OPE side through the relation,
\begin{equation}
m_{a_0}=\left[\frac{M^3}{2\hat{\Pi}^{\rm OPE}}\frac{\partial \hat{\Pi}^{\rm OPE}}{\partial M} \right]^{1/2}\label{ratio}\ .
\end{equation}
This way of extracting a hadronic mass is often adopted in QCD sum rules particularly for mesons.
In reality, however, since Eq.~(\ref{final}) is not exact by the rough assumption of QCD duality and the truncation in the OPE,
Eq.~(\ref{ratio}) may be limited in determining a hadron mass of concern precisely~\cite{Leinweber:1995fn}.
But our standpoint is that Eq.~(\ref{ratio}) is still useful as a qualitative guide in determining the possible structure of $a_0(980)$.

In our analysis, we take the continuum threshold corresponding to the mass of $a_0(1450)$, i.e., $s_0=(1.45~{\rm GeV})^2$.
$a_0(1450)$ is the next higher resonance with the same quantum numbers whose decay width is relatively large around 260 MeV.
So this choice seems to be consistent with the usual pole and continuum ansatz for the phenomenological side.
One worry though is that $a_0(1450)$ is the companion state of $a_0(980)$ related by the tetraquark mixing framework.
That is, its wave function is orthogonal to $a_0(980)$ in the constituent quark picture.
Then, one may wonder whether the interpolating field, Eq.~(\ref{if}), which was constructed optimally for $a_0(980)$,
does not couple to $a_0(1450)$ at all, denying its participation in the continuum.
But it should be remembered that the interpolating fields are composed of current quarks
so the features established in the constituent quark picture are not necessarily sustained in the current quark picture.
The coupling strength might be small but it is not zero.
In principle, a detail analysis might be necessary in determining the continuum threshold by scrutinizing the OPE
terms carefully~\cite{Leinweber:1995fn}.
But our simple prescription might be enough for our present purpose
in testing the reliability of the interpolating field like Eq.~(\ref{if}) for the light nonet.

In passing, it is also worth mentioning about some limitations in applying a QCD sum rule to
the heavy nonet member, $a_0(1450)$.
To construct a QCD sum rule for $a_0(1450)$, one can introduce an interpolating field for $a_0(1450)$
based on its static analog in Eq.~(\ref{heavy}), namely,
\begin{equation}
J^H_{a_0}=-\alpha J_{0}+\beta J_{1}\label{if2}\ ,
\end{equation}
and proceed the calculation similarly as above.
But this sum rule has a serious flaw from the fact that the interpolating field can couple to
both, $a_0(980), a_0(1450)$.
Even though the interpolating field, Eq.~(\ref{if2}), is optimal for $a_0(1450)$,
it still can couple to $a_0(980)$. Then the problem is that the unwanted resonance $a_0(980)$
constitutes the lowest-lying pole whose contribution is amplified
by the Borel weight in this $a_0(1450)$ sum rule.
Therefore, predictions from the $a_0(1450)$ sum rule are contaminated by the unwanted
lowest-lying pole contribution.

\section{Results and Discussion}
\label{sec:result}

We now present and discuss the results from the sum rule, Eq.~(\ref{final}), based on the OPE [Eq.~(\ref{ope})]
provided through Eq.~(\ref{ope00}),(\ref{ope01}),(\ref{ope11}).
Our discussion firstly focuses on
each QCD sum rule constructed from $J_0$ and $J_1$ separately.  Each sum rule will be examined in terms of its
reliability in predicting the $a_0(980)$ mass.
We then discuss the full sum rule constructed from $J^L_{a_0}$
which, through Eq.~(\ref{if}), is a linear combination of the two fields $J_0,J_1$.

\begin{figure} 
    \centerline{\epsfig{file=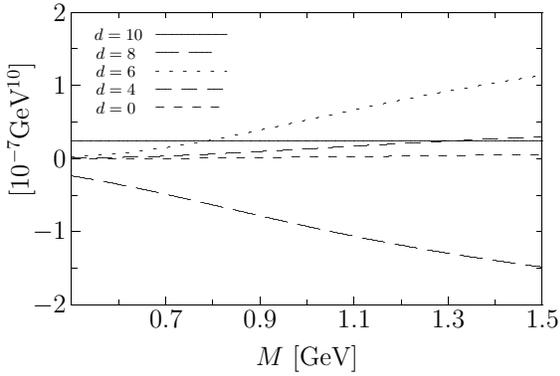,width=8cm,angle=0}}
    \caption{The Borel curves contributing to $\hat{\Pi}^{\rm OPE}_{0,0}$ [Eq.~(\ref{ope00})], plotted separately for
    each OPE dimension as specified in inset.}
    \label{scalar corr each}
\end{figure}

\begin{figure} 
    \centerline{\epsfig{file=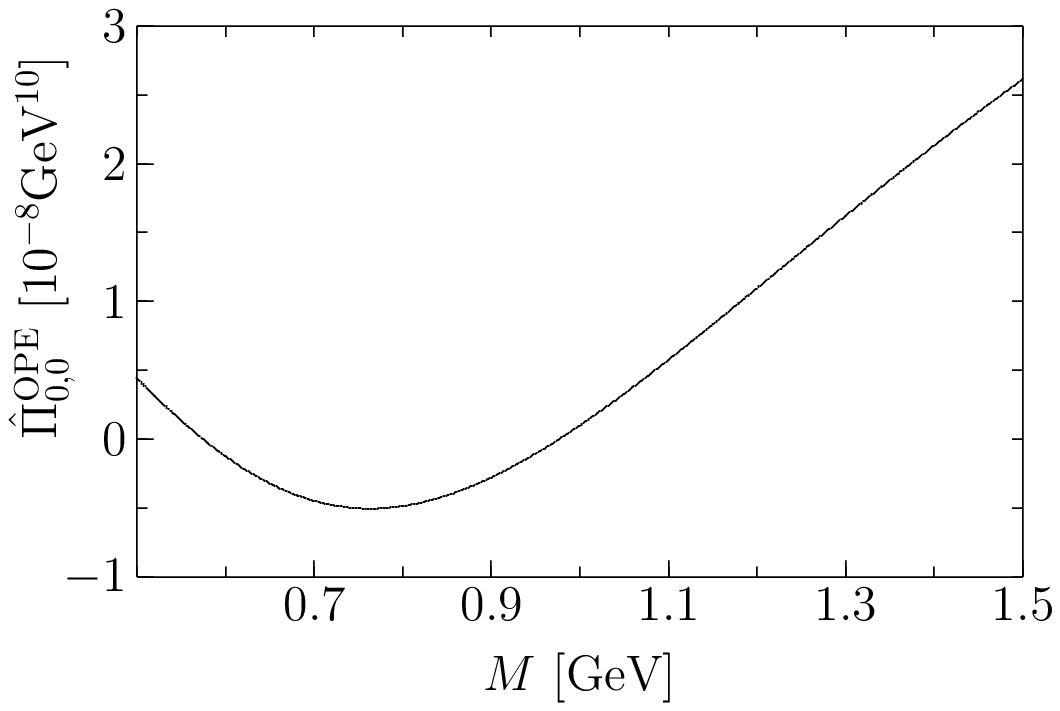,width=8.5cm,angle=0}}
    \caption{The Borel curve for $\hat{\Pi}^{\rm OPE}_{0,0}$, that is, the sum of all the lines in Fig.~\ref{scalar corr each}.}
    \label{scalar corr total}
\end{figure}

\begin{figure} 
    \centerline{\epsfig{file=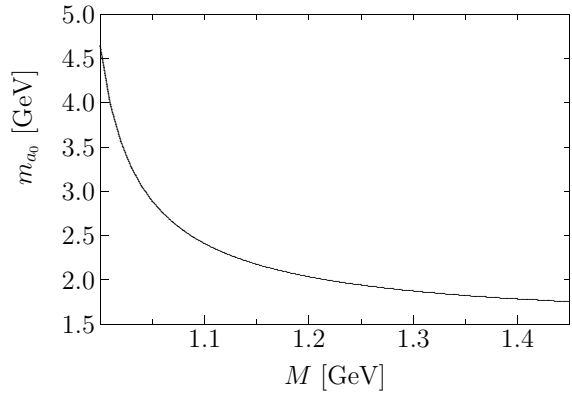,width=8cm,angle=0}}
    \caption{The Borel curve for the $a_0(980)$ mass extracted from Eq.~(\ref{ratio})
    using the interpolating field $J_0$ only.}
    \label{mass from scalar corr}
\end{figure}

We start with the QCD sum rule constructed only from the interpolating field $J_0$ [Eq.~(\ref{type0})] which
contains the spin-0 diquark.  This sum rule can be obtained from Eq.~(\ref{final}) [also see Eq.~(\ref{ope})]
by setting the mixing parameters $\alpha=0, \beta=1$. The OPE part in this case is, therefore,
given by $\hat{\Pi}^{\rm OPE}_{0,0}$ [Eq.~(\ref{ope00})].
From this sum rule, we reconfirm the result from Ref.~\cite{Lee:2005hs} that the interpolating field $J_0$
is not relevant for $a_0(980)$ in viewing the fact that the OPE is inconsistent with the left-hand side of Eq.~(\ref{final})
being positive definite~\footnote{Our calculation in this sum rule is slightly updated from Ref.~\cite{Lee:2005hs}
by including dimension 10 operators.}.
In particular, we plot in Fig.~\ref{scalar corr each} various contributions
to $\hat{\Pi}^{\rm OPE}_{0,0}$ classified according to the OPE dimension.
One can see that
the most important contribution comes from dimension 8 operators but its value is negative.
This negative value is driven mainly by the term containing
$\left[\langle\bar{q}q\rangle \langle\bar{s}g_c\sigma\cdot Gs\rangle
+\langle\bar{s}s\rangle \langle\bar{q}g_c\sigma\cdot Gq\rangle\right]$,
which even makes Eq.~(\ref{ope00}) totally negative
in the Borel region $0.5~\text{GeV} \le M\le 1.0~\text{GeV}$ (Fig.~\ref{scalar corr total}).
The curve in Fig.~\ref{scalar corr total} is not even similar in shape to the Borel curve roughly
plotted from the phenomenological form $\sim e^{-m^2_{a_0}/M^2}$ with $m_{a_0}\sim 1$ GeV.
Thus, the matching formula, Eq.~(\ref{final}), makes no sense in this case with $\alpha=0, \beta=1$.
In other words, the OPE side is simply incompatible with its phenomenological side.
Furthermore, the fact that the high dimensional operators of dimension 8 and 10 contribute dominantly to the OPE
already indicates that the truncation in the OPE is not appropriate in this sum rule.

Even so, one may blindly estimate the $a_0(980)$ mass from the Borel region $M\gtrsim 1$ GeV but
the result is not conclusive at all.
To show this, we plot in Fig.~\ref{mass from scalar corr} the $a_0(980)$ mass calculated from Eq.~(\ref{ratio})
with respect to the Borel mass in this case with $\alpha=0, \beta=1$.
The curve is very sensitive to the Borel mass so it seems almost impossible to choose any plateau from which one can
extract the $a_0(980)$ mass. All these results support that the sum rule with $J_0$ alone
simply fails in predicting the $a_0(980)$ mass.
Therefore, we conclude that the interpolating field $J_0$ [Eq.~(\ref{type0})] containing the spin-0 diquark configuration
only does not represent $a_0(980)$ properly.

\begin{figure} 
    \centerline{\epsfig{file=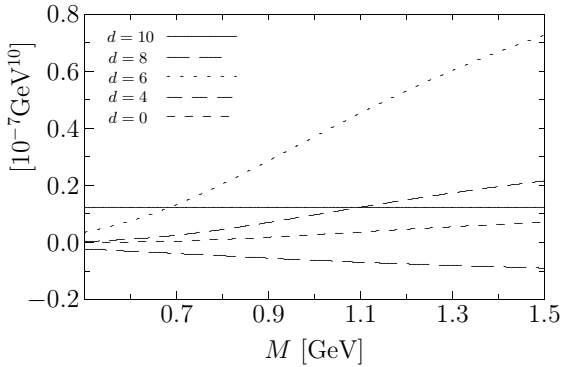,width=8cm,angle=0}}
    \caption{The Borel curves contributing to $\hat{\Pi}^{\rm OPE}_{1,1}$ [Eq.~(\ref{ope11})], plotted separately for
    each OPE dimension as specified in inset}
    \label{vector corr each}
\end{figure}

\begin{figure} 
    \centerline{\epsfig{file=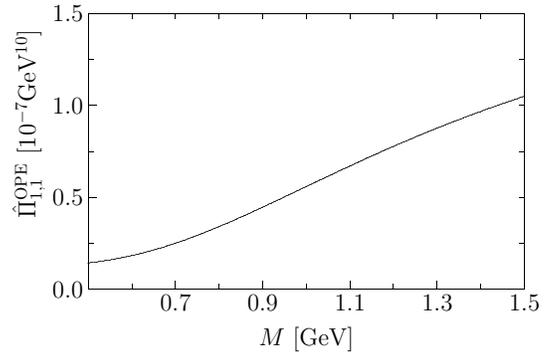,width=8cm,angle=0}}
    \caption{The Borel curve for $\hat{\Pi}^{\rm OPE}_{1,1}$, that is, the sum of all the lines in Fig.~\ref{vector corr each}.}
    \label{vector corr total}
\end{figure}

On the other hand, very different aspects can be observed from the sum rule using the interpolating field
$J_1$ [Eq.~(\ref{type1})] which contains the spin-1 diquark only.
This sum rule can be obtained by setting $\alpha=1,\beta=0$ in Eq.~(\ref{final}) [also see Eq.~(\ref{ope})] so its OPE side is given
by $\hat{\Pi}^{\rm OPE}_{1,1}$ [Eq.~(\ref{ope11})].
Firstly, as one can see in Fig.~\ref{vector corr each}, the contribution from dimension 8 operators, which was dominant
in $\hat{\Pi}^{\rm OPE}_{0,0}$, becomes small with its values being negative.
This is mainly due to the cancelation in the term
$\left[\langle\bar{q}q\rangle \langle\bar{s}g_c\sigma\cdot Gs\rangle
-\frac{3}{2}\langle\bar{s}s\rangle \langle\bar{q} g_c\sigma\cdot Gq\rangle\right]$ in Eq.~(\ref{ope11}).
The full OPE in this case is positive as it is mainly driven by dimension 6 operators
so that this sum rule at least is not
contradictory to the positive constraint imposed by the left-hand side of Eq.~(\ref{final}).
Secondly, the high dimensional operators at dimension 8 and 10 take up
a small portion in the OPE, which
qualitatively guarantees the OPE convergence in this calculation up to dimension 10.
Moreover, the Borel curve, if plotted for the full OPE (Fig.~\ref{vector corr total}) in this $J_1$ sum rule,
is similar in shape to the Borel curve roughly generated from the phenomenological form $\sim e^{-m^2_{a_0}/M^2}$.
So both sides can be matched at least qualitatively.
By comparing the vertical scales of Fig.~\ref{scalar corr total} and Fig.~\ref{vector corr total}, one can
see that $\hat{\Pi}^{\rm OPE}_{1,1}$ is much larger than $\hat{\Pi}^{\rm OPE}_{0,0}$ in most Borel region.
Thus, $\hat{\Pi}^{\rm OPE}_{1,1}$ must be an important part even in the final sum rule later
with the full interpolating field $J_{a_0}^L$ [Eq.~(\ref{if})].

\begin{figure} 
    \centerline{\epsfig{file=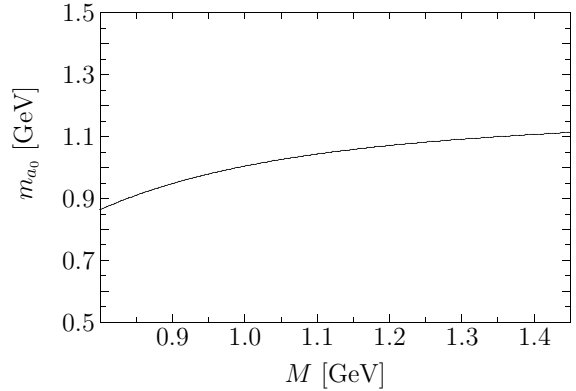,width=8cm,angle=0}}
    \caption{The Borel curve for the $a_0(980)$ mass extracted from Eq.~(\ref{ratio}) using the interpolating field $J_1$ only.}
    \label{mass from vector corr}
\end{figure}

We investigate this $J_1$ sum rule further by calculating the $a_0(980)$ mass from Eq.~(\ref{ratio}).
The Borel curve is drawn in Fig.~\ref{mass from vector corr}, which is much flatter than
the $J_0$ case in Fig.~\ref{mass from scalar corr}. The extracted mass is very good
but it still has moderate dependence on $M$.
Specifically, we see from the figure that the $a_0(980)$ mass
varies between $0.86~\text{GeV}<m_{a_0}<1.11~\text{GeV}$ within the Borel range $0.8~\text{GeV}<M<1.45~\text{GeV}$.
The middle value from the mass window, $m_{a_0}\sim 0.985$ GeV, agrees very well
with the experimental mass but the extraction error is somewhat large as $\Delta m_{a_0}\sim 0.25$ GeV.
One reason for this error can be traced to the fact that the OPE is dominated
by the dimension 6 contribution.
It is not difficult to see from Fig.~\ref{vector corr each} that
the other OPE terms, even if they are all summed up, are rather
small compared to the dimension 6 contribution.
Nevertheless, from all the nice aspects discussed above,
we can claim that the interpolating field with the
spin-1 diquark configuration is much more
promising in describing $a_0(980)$ than the one with the spin-0 configuration as far as the QCD sum rule
analysis is concerned.

\begin{figure} 
    \centerline{\epsfig{file=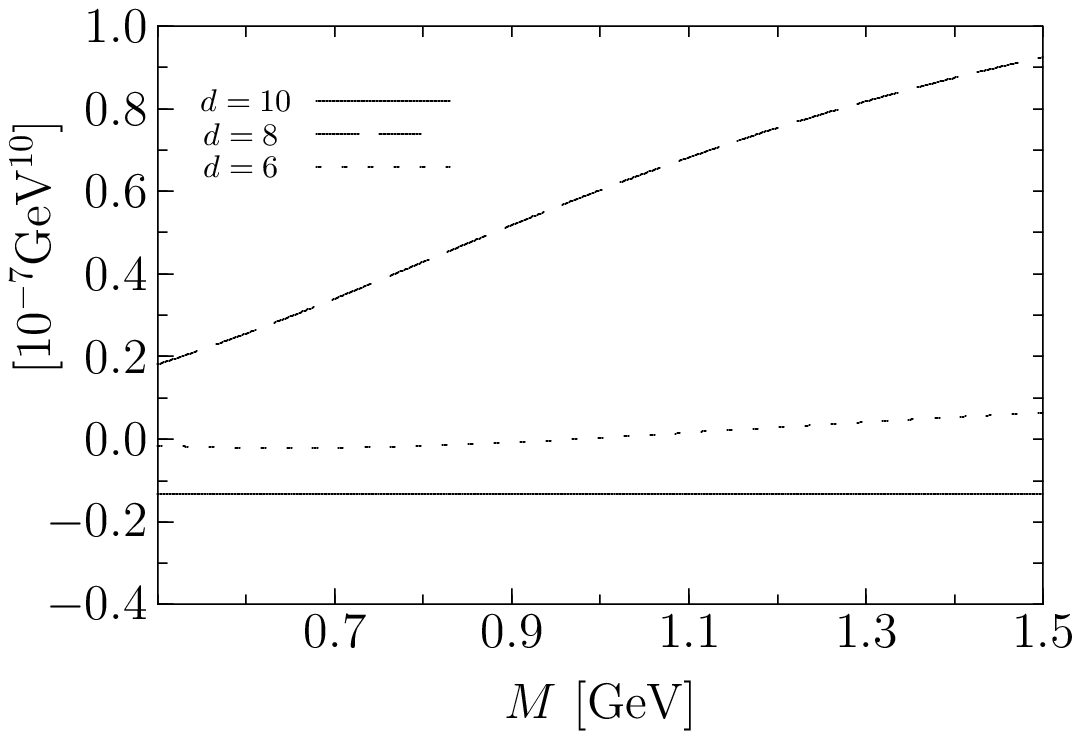,width=8cm,angle=0}}
    \caption{The Borel curves contributing to $\hat{\Pi}^{\rm OPE}_{0,1}$ [Eq.~(\ref{ope01})], plotted separately for
    each OPE dimension as specified in inset.}
    \label{mixed corr each}
\end{figure}

\begin{figure} 
    \centerline{\epsfig{file=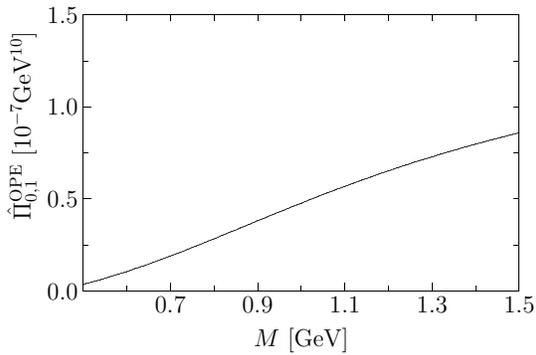,width=8cm,angle=0}}
    \caption{The Borel curve for the mixed correlator, $\hat{\Pi}^{\rm OPE}_{0,1}$ [Eq.~(\ref{ope01})].}
    \label{mixed corr total}
\end{figure}

There are another interesting results to discuss from the mixed correlator
whose OPE is given by $\hat{\Pi}^{\rm OPE}_{0,1}$ [Eq.~(\ref{ope01})].
First of all, this mixed correlator is different from $\hat{\Pi}^{\rm OPE}_{0,0}$, $\hat{\Pi}^{\rm OPE}_{1,1}$
in that this correlator alone has no phenomenological counterpart.
As one can see in Fig.~\ref{mixed corr each},
its OPE is dominated by dimension 8 operators.
Consequently, the full OPE from this mixed correlator grows as the Borel mass
increases (Fig.~\ref{mixed corr total}).
This shape is similar to Fig.~\ref{vector corr total} supposedly reinforcing the
trends from $\hat{\Pi}^{\rm OPE}_{1,1}$ in the full sum rule below.
The dimension 10 contribution is quite small
as one can see from the solid line in Fig.~\ref{mixed corr each}. Main observation to make
is that the total strength of $\hat{\Pi}^{\rm OPE}_{0,1}$ is fairly large in magnitude
when compared to the other correlation functions.
At $M\sim 1$ GeV, $\hat{\Pi}^{\rm OPE}_{0,1} = 4.8\times 10^{-8}$ GeV$^{10}$,
which is comparable to
$\hat{\Pi}^{\rm OPE}_{1,1} = 5.6\times 10^{-8}$ GeV$^{10}$ but it is much larger than any value of
$\hat{\Pi}^{\rm OPE}_{0,0}$ in Fig.~\ref{scalar corr total}.

We now discuss the final sum rule of Eq.~(\ref{final}) constructed from the full interpolating field
$J_{a_0}^L$ [Eq.~(\ref{if})] with the mixing parameters, $\alpha=0.8167$, $\beta=0.5770$ as determined by Ref.~\cite{Kim:2016dfq}.
Here, all the three correlation functions, $\hat{\Pi}^{\rm OPE}_{0,0}$, $\hat{\Pi}^{\rm OPE}_{0,1}$, $\hat{\Pi}^{\rm OPE}_{1,1}$,
participate in making the OPE side of Eq.~(\ref{final}).
They are combined through Eq.~(\ref{ope}) so that the relative contribution from each correlator is subject to
further modulation by the fact that $\alpha > \beta$.  This modulation turns out to be encouraging in two respects.
First, the correlator with $J_0$, that is $\hat{\Pi}^{\rm OPE}_{0,0}$, is multiplied by $\beta^2$
so its contribution
is suppressed relatively more than those from $\hat{\Pi}^{\rm OPE}_{0,1}$ and $\hat{\Pi}^{\rm OPE}_{1,1}$.
This suppression is very nice because $\hat{\Pi}^{\rm OPE}_{0,0}$,
whose sum rule contains the unpleasant features as discussed above,
becomes less important in the final sum rule.
Second, the $\hat{\Pi}^{\rm OPE}_{1,1}$ contribution is enhanced by the parameter $\alpha^2$.
This can be regarded as another encouraging point because the sum rule with $J_1$ alone, which
already has various nice features as discussed above, becomes more important in the final sum rule.

The contribution from the mixed correlator, $\hat{\Pi}^{\rm OPE}_{0,1}$, is also amplified by the
factor $2\alpha\beta$ in Eq.~(\ref{ope}),
which is about 40\% larger than the $\alpha^2$ factor.
Recalling that $\hat{\Pi}^{\rm OPE}_{0,1}$ is slightly less than $\hat{\Pi}^{\rm OPE}_{1,1}$,
the modulated contribution from the mixed correlator in Eq.~(\ref{ope})
becomes even larger than that from $\hat{\Pi}^{\rm OPE}_{1,1}$.
This indicates that this mixed correlator constitutes an important part in the final sum rule, Eq.~(\ref{final}).
This finding seems to be consistent with the tetraquark mixing framework
established in the constituent quark picture where the two tetraquark types mix strongly through the
color-spin interaction~\cite{Kim:2016dfq,Kim:2017yur,Kim:2017yvd,Kim:2018zob}.
Although a direct connection between the two approaches needs to be clarified, this consistency
could be another evidence to support the tetraquark mixing framework.

\begin{figure} 
    \centerline{\epsfig{file=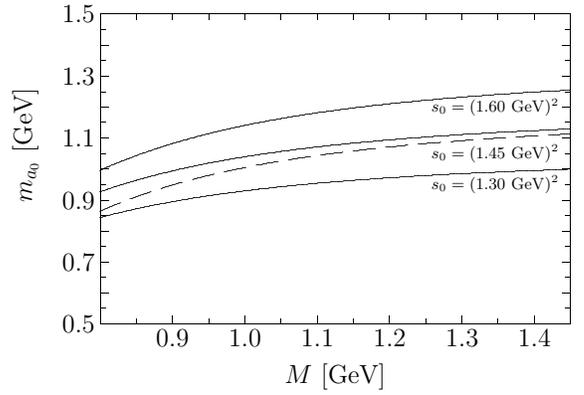,width=8cm,angle=0}}
    \caption{The Borel curves for the $a_0(980)$ mass extracted from Eq.~(\ref{ratio}) using the full
    interpolating field $J_{a_0}^L$ [Eq.~(\ref{if})] are presented here with three solid lines with different $s_0$ as specified.
    The dashed line is the same curve as in Fig.~\ref{mass from vector corr} plotted here again for a clear comparison.}
    \label{mass from full corr}
\end{figure}

In Fig.~\ref{mass from full corr}, we plot the Borel curves for $m_{a_0}$ calculated
from Eq.~(\ref{ratio}) using the full OPE in Eq.~(\ref{ope}). Our result shown by
the solid line in the middle is obtained by using the same continuum threshold
$s_0=(1.45~{\rm GeV})^2$ as the dashed line [the same curve in Fig.~\ref{mass from vector corr}]
calculated only from $\hat{\Pi}^{\rm OPE}_{1,1}$.
Focusing only on this result for the time being, we see that
the extracted mass varies from $0.93~\text{GeV}<m_{a_0}<1.13~\text{GeV}$
so its middle value is 1.03 GeV
with the extraction error $\Delta m_{a_0} \sim 0.2$ GeV.
This Borel curve is slightly flatter than the one in Fig.~\ref{mass from vector corr}.
Thus, $m_{a_0}$ from this full sum rule is not so different from the one from $\hat{\Pi}^{\rm OPE}_{1,1}$
and also from the experimental mass.
The large contribution from the mixed correlator, $\hat{\Pi}^{\rm OPE}_{0,1}$, seems to
cancel away mostly through the ratio, Eq.~(\ref{ratio}), so the extraction of $m_{a_0}$
turns out not to depend much on $\hat{\Pi}^{\rm OPE}_{0,1}$.

This result shows a physical role of the mixed correlator
when we see it in the context of Eq.~(\ref{final}).
Its large contribution certainly increases the right-hand side of Eq.~(\ref{final}) substantially
but, as we have shown, its inclusion does not change much the extraction of $m_{a_0}$
in the left-hand side. Then, its contribution must participate in increasing the other
hadronic parameter in Eq.~(\ref{final}), the coupling strength $f_{a_0}$.
This implies that the interpolating field $J^L_{a_0}$, through Eq.~(\ref{coup}),
couples to the lowest-lying resonance $a_0(980)$ more strongly when it is represented by the mixture
of the form, Eq.~(\ref{if}).

As we have mentioned earlier, predictions from QCD sum rules
may suffer from various uncertainties coming from
the rough prescriptions adopted.
Among various prescriptions, the major uncertainty in our sum rules
comes from the continuum threshold, $s_0$.
To demonstrate this, we plot the Borel curve with the larger threshold $s_0 = (1.60~{\rm GeV})^2$
as shown in the upper solid curve in Fig.~\ref{mass from full corr}.
The middle value between the upper and lower mass bounds of the curve
becomes 1.13 GeV. This value is 10\% larger than the above value of 1.03 GeV.
Similarly, the smaller threshold $s_0=(1.30~{\rm GeV})^2$ produces the lower solid curve,
which yields the middle mass 0.92 GeV, i.e., 10\% smaller than 1.03 GeV.
This 10\% error is more or less endurable in QCD sum rules if one
considers the abrupt nature in treating the continuum as explained in Sec.~\ref{sec:QCDSR}.
Anyway, the fact that our extracted mass is around 1 GeV even if we take into account the uncertainty in $s_0$,
can be used as a qualitative guide in supporting that
the interpolating field in Eq.~(\ref{if}) is relevant for $a_0(980)$.

The vacuum saturation hypothesis in factorizing high dimensional operators could be another
source of uncertainty in our sum rules
because the operators in dimension 6,8,10, which have been estimated by the vacuum saturation hypothesis,
are the important part of the OPE.
As discussed in Refs.~\cite{Shifman:1978bx,Novikov:1983jt},
the vacuum saturation hypothesis is justified by the $1/N_c$ expansion and
its correction can be estimated by inserting other intermediate states.
The deviation from this hypothesis is expected to be around 10\%~\cite{Reinders:1984sr}
although there are some reports with bigger deviations as described in Ref.~\cite{Leinweber:1995fn}.
So, to estimate the uncertainty from this assumption, we increase 10\% for the high dimensional operators
that have been factorized in Eqs.~(\ref{ope00}),(\ref{ope01}),(\ref{ope11}).  Our numerical calculation shows
that the extracted mass is 1.02 GeV, only 1\% smaller than the factorized result, 1.03 GeV.
The other uncertainties from the truncation in the OPE are not so
important in our sum rules as the dimension 10 operators are
already small enough in the full sum rule above.

But, instead of dwelling on a reliability of the mass prediction,
what is more important to us is the fact that our sum rule delivers three solid
statements related to the structure of $a_0(980)$, which are not affected much
by the rough prescriptions.
The first statement is that the interpolating field $J_1$, which
involves the spin-1 diquark configuration only, is the main driving force in
producing the sum rule result.
As we have discussed, the sum rule with $J_1$ alone already has various nice features.
Only the exception is the fact that
the Borel curve is not flat enough to pin down a certain mass for $a_0(980)$.
The second statement is that the mixed correlation function $\hat{\Pi}^{\rm OPE}_{0,1}$ also contributes
appreciably to the sum rule. This is essentially consistent with what the tetraquark mixing framework is advocating.
Its role is to strengthen the coupling $f_{a_0}$ which represents
an overlap of the interpolating field $J^L_{a_0}$ with the physical $a_0(980)$.
The third statement is that the correlator, $\hat{\Pi}^{\rm OPE}_{0,0}$, which is constructed
from the $J_0$ interpolating field only, contributes minimally to the $a_0(980)$ sum rule.
This last statement is very different from the common expectation that $a_0(980)$
is a tetraquark mostly with the spin-0 diquark configuration.
But this does not mean that spin-0 diquark configuration is totally irrelevant to describe $a_0(980)$.
This configuration contributes to our sum rule through the mixed
correlator, $\hat{\Pi}^{\rm OPE}_{0,1}$, that constitutes another important ingredient in our sum rule
as we have already mentioned.

To conclude, our sum rule supports that the interpolating field Eq.~(\ref{if}),
whose form is motivated by the tetraquark mixing framework, represents $a_0(980)$ reasonably well.
Our final sum rule has the moderate Borel stability and the OPE convergence.
From a detail analysis of the sum rule, we demonstrate that the tetraquark structure of $a_0(980)$ is
dominated by the spin-1 diquark configuration and its mixing with the spin-0 diquark configuration.
But the QCD sum rule constructed from the spin-0 diquark configuration alone fails to predict the $a_0(980)$ mass.
Our results therefore support the tetraquark mixing framework for the two light-meson nonets
established in the constituent quark picture~\cite{Kim:2016dfq,Kim:2017yur,Kim:2017yvd,Kim:2018zob}.

Our analysis has been performed only for the isovector resonance $a_0(980)$ in this work.
This analysis can be extended trivially to the isodoublet member
$K^*_0(800)$ in the light nonet because $a_0(980)$ and $K^*_0(800)$
are simply related by the SU(3)$_f$ symmetry when they are viewed from the tetraquark mixing framework.
The SU(3)$_f$ breaking, which is governed by the strange-quark mass in this case,
contributes marginally to the sum rule.
This means, the spin-1 diquark configuration
and its mixing should be also important to explain $K^*_0(800)$ with similar characteristics.
But the situation can be nontrivial for the isoscalar resonances $f_0(500)$, $f_0(980)$ due to flavor mixing.
The resonances $f_0(500)$ and $f_0(980)$ are not the definite flavor members of the octet and the singlet.
Instead, they are the mixtures of the two multiplets according to the generalized OZI rule~\cite{Kim:2017yvd}.
In future, it will be interesting to investigate
the role of the spin-1 diquark configuration in these resonances using QCD sum rules.

\section{Summary}
\label{sec:summary}

To summarize, we have performed in this work a QCD sum rule analysis
for $a_0(980)$ based on the tetraquark mixing framework
recently proposed in order to explain the two light-meson nonets.
Motivated by the mixing framework, we construct an interpolating field for $a_0(980)$ which
can reproduce the spin-0 and spin-1 diquark configurations in the static limit.
We then constructed QCD sum rules for $a_0(980)$ by calculating the OPE up to dimension 10 operators.
The OPE expression is divided into three correlation functions depending on the participating
interpolating fields.
The first correlator is composed of the interpolating fields with the spin-0 diquark configuration only,
the second correlator with the spin-1 diquark configuration only, and the third correlator the
mixed type of the two configurations.
We have performed a detail analysis to identify the role of each correlation function in the sum rule.
We found that the spin-1 diquark configuration is very important to generate the $a_0(980)$ mass and the
mixed correlator also constitutes an important part in the total OPE.  The first correlator
only with the spin-0 diquark configuration contributes to the final sum rule marginally.
The last point is quite different from the common expectation that the $a_0(980)$ is a tetraquark
containing the spin-0 diquark configuration.
This work may help in establishing an interesting view on the tetraquark structure of $a_0(980)$,
that is, the state containing the spin-1 diquark configuration as well as the spin-0 diquark configuration.

\acknowledgments

The work of H.-J. Lee was supported by the Basic Science Research Program through the National Research
Foundation of Korea (NRF) funded by Ministry of Education under Grant No. 2016R1D1A1A09920078.
The work of H. Kim and K.S.Kim was supported by the National Research Foundation of Korea(NRF) grant funded by the
Korea government(MSIT) (No. NRF-2018R1A2B6002432 and No. NRF-2018R1A5A1025563).

\end{document}